\documentclass[useAMS,usenatbib]{mn2e} \usepackage{amsmath,amssymb}
\usepackage{graphicx,mathptm} \usepackage{times} \usepackage{natbib}
\usepackage{psfig}
\newcommand{\Mpch}{$h^{-1}\,\mbox{Mpc}$\,}
\newcommand{\xii}{$\xi(r_\perp,r_\parallel)$\,}
\newcommand{\xiis}{$\xi(s_\perp,s_\parallel)$\,}

\newcommand{\eg}{{e.g.}}

\title[Redshift-space distortions in cDE models] {Clustering and
  redshift-space distortions in interacting dark energy cosmologies}

\author[Marulli, Baldi \& Moscardini] {Federico Marulli$^{1,2,3}$,
  Marco Baldi$^{4,5}$ and Lauro
  Moscardini$^{1,2,3}$\\ $^1$Dipartimento di Astronomia, Alma Mater
  Studiorum - Universit\`a di Bologna, via Ranzani 1, I-40127 Bologna,
  Italy\\ $^2$INAF-Osservatorio Astronomico di Bologna, Via Ranzani 1,
  40127, Bologna, Italy\\ $^3$INFN/National Institute for Nuclear
  Physics, Sezione di Bologna, viale Berti Pichat 6/2, I-40127
  Bologna, Italy\\ $^4$ Excellence Cluster Universe, Boltzmannstr. 2,
  D-85748 Garching, Germany\\ $^5$ University Observatory,
  Ludwig-Maximillians University Munich, Scheinerstr. 1, D-81679
  Munich, Germany\\
}

\begin{document}

\maketitle

\begin{abstract}
We investigate the spatial properties of the large scale structure
(LSS) of the Universe in the framework of coupled dark energy (cDE)
cosmologies. Using the public halo catalogues from the 
{\small CoDECS} simulations -- the largest set of
N-body experiments to date for such cosmological scenarios -- 
we estimate the clustering and bias
functions of cold dark matter (CDM) haloes, both in real- and
redshift-space. Moreover, we investigate the effects of the dark
energy (DE) coupling on the geometric and dynamic redshift-space
distortions, quantifying the difference with respect to the
concordance $\Lambda $CDM model. At $z\sim 0$, the spatial properties
of CDM haloes in cDE models appear very similar to the $\Lambda $CDM
case, even if the cDE models are normalized at last scattering in
order to be consistent with the latest Cosmic Microwave Background
(CMB) data. At higher redshifts, we find that the DE coupling produces
a significant scale-dependent suppression of the halo clustering and
bias function. This effect, that strongly depends on the coupling
strength, is not degenerate with $\sigma_{8}$ at scales $r\lesssim 5-10$
\Mpch. Moreover, we find that the coupled DE strongly affects both the
linear distortion parameter, $\beta$, and the pairwise peculiar
velocity dispersion, $\sigma_{12}$. Although the models considered in
this work are found to be all in agreement with presently available
observational data, the next generation of galaxy surveys will be able
to put strong constraints on the level of coupling between DE and CDM
exploiting the shape of redshift-space clustering anisotropies.
\end{abstract}



\section {Introduction}
\label{intro}
The concordance $\Lambda $CDM cosmological model has strongly improved
its reputation over the last two decades thanks to the overall
agreement between its predictions and the ever increasing wealth of
available observational data \citep[see e.g.][]{riess1998,
  perlmutter1999, percival2001, spergel2003, astier2006, komatsu2010,
  crocce2011, carnero2011}.  However, some possible tensions between
the predictions of the $\Lambda $CDM scenario and astrophysical
observations at small scales have also been claimed, like the
abundance of satellite galaxies in CDM haloes, the low baryon fraction
in galaxy clusters and the `cusp-core' problem for the halo density
profiles \citep[see e.g.][]{allen2004, laroque2006, vikhlinin2006,
  simon2003, binney2001, newman2009}. It is still unclear whether such
discrepancies might be due to an incorrect description of the baryonic
phenomena at work, or if they are directly connected to the underlying
cosmological scenario. In any case, it is interesting to investigate
how the standard $\Lambda $CDM model could be modified in order to
better match the above-mentioned observations.

Furthermore, the fundamental nature of CDM particles and of the DE
component -- that together constitute more than 95\% of the total
energy of the Universe -- is still unknown, which represents one of
the central problems of modern cosmology.  From an observational point
of view, the existence of a stress-energy component with negative
pressure, responsible for the accelerated expansion of the Universe,
can be directly deduced from supernova measurements \citep[see
  \eg][]{riess1998,
  perlmutter1999,Schmidt_etal_1998,SNLS,Kowalski_etal_2008}, or by
jointly combining the CMB power spectrum and low redshift
observations, like measurements of the Hubble constant, of the galaxy
power spectrum and of cluster counts. More recently, by combining the
CMB lensing deflection power spectrum from the Atacama Cosmology
Telescope with temperature and polarization power spectra from the
Wilkinson Microwave Anisotropy Probe, it has been possible to break
the geometric degeneracy and constrain the dark energy density using
measurements of the CMB radiation alone \citep{sherwin2011}.

Several theoretical efforts have been made in order to find a
satisfactory explanation for the cosmic acceleration. Different DE
models have been proposed in the literature, ranging from simple
scenarios with a constant equation of state to models considering
interactions between DE and other cosmic fluids, as e.g. CDM
\citep[see e.g.][]{wetterich1995, amendola2000} or massive neutrinos
\citep[][]{amendola2008}.  Indeed, if the accelerated expansion is
driven by a scalar field, there is no fundamental reason why DE and
CDM should not interact with each other.  Interestingly, it has been
shown that such a direct coupling could help in alleviating some of
the small-scale problems of the $\Lambda $CDM cosmology \citep[see
  e.g.][]{Baldi_etal_2010,Baldi_Pettorino_2011,Lee_Baldi_2011}.

In the present paper, we investigate the spatial properties of LSS
predicted by interacting DE cosmological models. In particular, this
work is focused on the clustering and bias properties of CDM haloes
and on their redshift-space clustering distortions (RSD) caused by the
line-of-sight component of galaxy peculiar velocities. The latest
galaxy surveys that have enabled measurements of RSD are the 2-degree
Field Galaxy Redshift Survey \citep{peacock2001, hawkins2003,
  percival2004}, the Sloan Digital Sky Survey \citep{tegmark2004,
  zehavi2005, tegmark2006, okumura2008, cabre2009, cabre2009b} and the
VIMOS-VLT Deep Survey \citep{guzzo2008}. 

RSD can be used to probe gravity theories, as they are directly
related to the growth rate of LSS \citep{linder2008}. Moreover, they
can also be used in several other contexts, e.g. to robustly constrain
the value of the total mass of cosmological neutrinos
\citep{marulli2011} or to investigate the dynamical properties of the
warm-hot intergalactic medium \citep{ursino2011}. Since the
interaction between DE and CDM strongly affects the formation and
evolution of LSS, a significant effect on RSD is also expected. To
properly describe all the non-linear effects at work, we use a series
of state-of-the-art N-body simulations for a wide range of different
cDE scenarios -- the {\small CoDECS} simulations \citep{CoDECS} -- and
discuss the impact that DE interactions can have on LSS with respect
to the $\Lambda $CDM case.  Both on-going and next generation galaxy
surveys, such as VIPERS (Guzzo et al. 2011 in preparation), BigBOSS
\citep{schlegel2009} and Euclid \citep{laureijs2009,EUCLID-y}, will
greatly improve the precision in RSD measurements over an even grater
redshift range, up to $z=2$, allowing for direct constraints on the
underlying cosmological model and on the gravity theory.

The structure of the paper is as follows. In section \S \ref{cDE} we
describe the cDE models analyzed in this work, while in section \S
\ref{N-body} we introduce the exploited set of N-body experiments used
to simulate the LSS in these cosmological frameworks. In section \S
\ref{LSS} we analyze the CDM halo clustering and bias functions and
the redshift-space clustering distortions. Finally, in section \S
\ref{concl} we draw our conclusions.


\section {The Coupled Dark Energy models}
\label{cDE}

Interacting DE scenarios have been introduced as a possible
alternative to the standard $\Lambda $CDM cosmology
\citep{wetterich1995,amendola2000} with the aim of addressing the
fine-tuning problems that characterize the cosmological
constant. These alternative cosmologies include models where a DE
scalar field $\phi$ interacts with other cosmic fluids by exchanging
energy-momentum during the evolution of the universe.  Several
possible forms of DE interaction have been proposed in the literature,
including e.g. models of coupling to the CDM fluid \citep[as
  \eg][]{amendola2000,amendola2004,
  Koyama_etal_2009,Honorez_etal_2010,Baldi_2011a} or to massive
neutrinos \citep[][]{amendola2008}.  In the present work, we will
focus on the former class, considering a DE-CDM interaction defined by
the following set of background dynamic equations:
\begin{eqnarray}
\label{klein_gordon}
\ddot{\phi } + 3H\dot{\phi } +\frac{dV}{d\phi } &=&
\sqrt{\frac{2}{3}}\eta _{c}(\phi ) \frac{\rho _{c}}{M_{{\rm Pl}}} \,,
\\
\label{continuity_cdm}
\dot{\rho }_{c} + 3H\rho _{c} &=& -\sqrt{\frac{2}{3}}\eta _{c}(\phi
)\frac{\rho _{c}\dot{\phi }}{M_{{\rm Pl}}} \,, \\
\label{continuity_baryons}
\dot{\rho }_{b} + 3H\rho _{b} &=& 0 \,, \\
\label{continuity_radiation}
\dot{\rho }_{r} + 4H\rho _{r} &=& 0\,, \\
\label{friedmann}
3H^{2} &=& \frac{1}{M_{{\rm Pl}}^{2}}\left( \rho _{r} + \rho _{c} +
\rho _{b} + \rho _{\phi} \right)\,,
\end{eqnarray}
where an overdot represents a derivative with respect to the cosmic
time $t$, $H\equiv \dot{a}/a$ is the Hubble function, $V(\phi)$ is the
scalar field self-interaction potential, $M_{\rm Pl}\equiv
1/\sqrt{8\pi G}$ is the reduced Planck Mass, and the subscripts
$b\,,c\,,r$ indicate baryons, CDM, and radiation, respectively. The
strength of the interaction between the DE scalar field and CDM
particles is defined by the coupling function $\eta_{c}(\phi)$, while
the shape of the potential $V(\phi)$ dictates the dynamical evolution
of the DE field that in turn determines a time evolution of the CDM
particle mass:
\begin{equation}
\label{mass}
\frac{d \ln M_{c}}{dt} = -\sqrt{\frac{2}{3}}\eta _{c}(\phi )\dot{\phi
}\,,
\end{equation}
as one can obtain from Eq.~(\ref{continuity_cdm}).

In this work, we will consider the set of coupled DE scenarios defined
in \citet{CoDECS}, which includes two possible choices for the
coupling function -- namely a constant coupling $\eta_{c} = \eta
_{0}$ and an exponential coupling $\eta_{c}(\phi )=\eta
_{0}\exp\left[\eta_{1}\phi\right] $-- and two possible forms of the
scalar potential, namely an exponential potential
\citep{lucchin1985,wetterich1988}:
\begin{equation}
\label{exponential}
V(\phi) = Ae^{-\alpha \phi}
\end{equation}
and a SUGRA potential \citep{brax1999}:
\begin{equation}
\label{SUGRA}
V(\phi) = A\phi ^{-\alpha }e^{\phi ^{2}/2} \,,
\end{equation}
where for simplicity the field $\phi $ has been expressed in units of
the reduced Planck mass in Eqs.~(\ref{exponential},\ref{SUGRA}). The
parameters of the models investigated in the present work are
summarized in Table~\ref{tab:models}.

\begin{table*}
\begin{center}
\caption{The list of cosmological models considered in the {\small
    CoDECS} project and their specific parameters.}
\begin{tabular}{llccccc}
\hline
\hline
Model & Potential & $\alpha$ & $\eta_{0}$ & $\eta_{1}$ & $w_{\phi}(z=0)$ & $\sigma _{8}(z=0)$ \\
\hline 
$\Lambda $CDM & $V(\phi ) =
A$ & -- & -- & -- & $-1.0$ & $0.809$ \\ 
EXP001 & $V(\phi ) =
Ae^{-\alpha \phi }$ & 0.08 & 0.05 & 0 & $-0.997$ & $0.825$ \\ 
EXP002 & $V(\phi ) = Ae^{-\alpha \phi }$ & 0.08 & 0.1 & 0 & $-0.995$ & $0.875$ \\ 
EXP003 & $V(\phi ) = Ae^{-\alpha \phi }$ & 0.08 & 0.15 & 0 & $-0.992$ & $0.967$ \\ 
EXP008e3 & $V(\phi ) = Ae^{-\alpha \phi }$ & 0.08 & 0.4 & 3 & $-0.982$ & $0.895$ \\ 
SUGRA003 & $V(\phi ) = A\phi^{-\alpha }e^{\phi ^{2}/2}$ & 2.15 & -0.15 & 0 & $-0.901$ & $0.806$ \\ 
\hline
\hline
\end{tabular}
\label{tab:models}
\end{center}
\end{table*}
\ \\

At the linear perturbations level, the time evolution of density
fluctuations in the coupled CDM fluid and in the uncoupled baryonic
component ($\delta _{c,b}\equiv \delta \rho _{c,b}/\rho _{c,b}$,
respectively) obeys the following system of equations:
\begin{eqnarray}
\label{gf_c}
\ddot{\delta }_{c} &=& -2H\left[ 1 - \eta _{c}\frac{\dot{\phi
  }}{H\sqrt{6}}\right] \dot{\delta }_{c} + 4\pi G \left[ \rho
  _{b}\delta _{b} + \rho _{c}\delta _{c}\Gamma _{c}\right] \,, \\
\label{gf_b}
\ddot{\delta }_{b} &=& - 2H \dot{\delta }_{b} + 4\pi G \left[ \rho
  _{b}\delta _{b} + \rho _{c}\delta _{c}\right]\,,
\end{eqnarray}
where for simplicity the field dependence of the coupling function
$\eta _{c}(\phi )$ has been omitted.  In Eq.~(\ref{gf_c}), the factor
$\Gamma _{c}\equiv 1 + 4\eta _{c}^{2}(\phi )/3$ encodes the effect of
the additional fifth-force mediated by the DE scalar field $\phi $ for
CDM perturbations. Additionally, the second term in the first square
bracket at the right-hand-side of Eq.~(\ref{gf_c}) represents an
extra-friction term on CDM fluctuations arising as a consequence of
momentum conservation \citep[see e.g.][for a derivation of
  Eqs.~(\ref{klein_gordon}-\ref{friedmann},\ref{gf_c},\ref{gf_b}) and
  for a detailed discussion of the extra-friction and fifth-force
  corrections to the evolution of linear perturbations]{amendola2004,
  pettorino2008,Baldi_etal_2010,Baldi_2011b}.

The background and linear perturbations evolution of all the
cosmologies considered in this work have been discussed in full detail
in \citet{Baldi_2011c} and \citet{CoDECS}, to which we refer the
interested reader for a more thorough description of the models. In
particular, the EXP008e3 model -- based on an exponential potential
and on an exponential coupling function -- features a steep growth of
the interaction strength at low redshifts and has been shown to
significantly enhance the pairwise infall velocity of colliding galaxy
clusters similar to the ``Bullet Cluster" \citep{Lee_Baldi_2011}.  On
the other hand, the SUGRA003 model based on a SUGRA potential and on a
negative constant coupling is an example of the recently proposed
``Bouncing" cDE scenario \citep{Baldi_2011c}, which is characterized
by an inversion of the scalar field direction of motion at relatively
recent epochs ($z_{\rm inv}\approx 6.8$ for the specific model under
investigation), giving rise to a ``bounce" of the DE equation of state
parameter $w_{\phi }$ on the cosmological constant barrier $w_{\phi
}=-1$.  As a consequence of such inversion, the quantity $\eta
_{c}\dot{\phi }$ that appears both in Eqs.~(\ref{mass}) and
(\ref{gf_c}) changes sign in correspondence to the bounce such that
the CDM particle mass decreases before the bounce and increases again
afterwards, while the extra-friction term changes from a drag to a
proper friction at $z_{\rm inv}$. Such peculiar dynamical behaviour
has been shown to provide a possible explanation \citep{Baldi_2011c}
to the recent detections of massive clusters of galaxies at high
redshifts \citep{jee2009,rosati2009}, although the statistical
significance of such detections as a challenge to the standard
$\Lambda $CDM scenario is presently not yet conclusive \citep[see
  e.g.][]{mortonson2011,waizmann2011}.


\section{The N-body simulations}
\label{N-body}

For our investigation we rely on the public halo catalogues of the
{\small CoDECS} simulations \citep{CoDECS}, the largest N-body
simulations for cDE cosmologies to date, that include all the models
described in Table~\ref{tab:models}. In particular, we make use of the
{\small L-CoDECS} suite which consists of large scale, collisionless
N-body runs following the evolution of $1024^{3}$ CDM particles and
$1024^{3}$ baryonic particles in a cosmological box of $1$ comoving
Gpc$/h$ on a side. The simulations have been carried out with the modified
version by \citet{Baldi_etal_2010} of the widely used parallel Tree-PM
N-body code {\small GADGET} \citep{springel2005gadget2}, and have a
mass resolution of $m_{c}(z=0) = 5.84\times 10^{10}$ M$_{\odot }/h$
and $m_{b} = 1.17\times 10^{10}$ M$_{\odot }/h$ for CDM and baryons,
respectively, and a force resolution of $\epsilon_{g} = 20$ kpc$/h$.

The initial conditions for all the different cosmologies were
generated by perturbing a homogeneous {\em ``glass"} distribution
\citep{white1994,baugh1995} based on the same linear power spectrum at
$z_{\rm CMB}$, and then rescaling the resulting displacements to the
starting redshift of the simulations, $z_{i} = 99$, with the specific
growth factor $D_{+}(z)$ of each model computed by numerically solving
Eqs.~(\ref{gf_c},\ref{gf_b}). This procedure ensures that all the
models are consistent with the same normalization of density
perturbations at the last scattering surface, which has been assumed
to correspond to the ``WMAP7 Only Maximum Likelihood" constraints of
\citet{komatsu2011}. The cosmological parameters at $z=0$ assumed for
all the models included in the {\small CoDECS} project are
$H_0=70.3\rm\,km\,s^{-1}\,Mpc^{-1}$, $\Omega_{\rm CDM}=0.226$,
$\Omega_{\rm DE}=0.729$, $\sigma_{8}=0.809$, $\Omega_b=0.0451$ and
$n_s=0.966$.

The halo catalogues have been obtained by identifying groups of
particles by means of a Friends-of-Friends \citep[FoF,][]{davis1985}
algorithm with linking length $\lambda = 0.2\times \bar{d}$, where
$\bar{d}$ is the mean interparticle separation. The FoF algorithm was
run over the CDM particles as primary tracers of the matter
distribution, and then baryonic particles have been attached to the
FoF group of their closest CDM neighbour. In order to identify CDM
substructures we have used the SUBFIND algorithm described in
\citet{springel2001}. All the results presented in this paper have
been obtained using mass-selected sub-halo catalogues, composed by the
gravitationally bound substructures that SUBFIND identifies in each
FoF halo. For all the simulations, we have restricted our analysis in
the mass range $M_{\rm min}<M<M_{\rm max}$, where $M_{\rm
  min}=2.5\cdot10^{12} M_\odot/h$ and $M_{\rm
  max}=3.6\cdot10^{15},1.1\cdot10^{15},4.9\cdot10^{14},2.6\cdot10^{14},1.8\cdot10^{14}
M_\odot/h$ at $z=0,0.55,1,1.6,2$, respectively.  For the analysis
presented in this paper, we have used one realisation of each
model. To investigate how reliable our results are, we have divided
each simulation box in 27 independent sub-boxes and estimated the
uncertainties in our predictions from the scatter between them. At the
scales considered, we found a very good agreement between the errors
estimated in this way and the theoretical ones, both for the
correlation function and for the bias.


\section {LSS in cDE models}
\label{LSS} 

In this section we show how the halo clustering, the bias function and
the redshift-space distortion parameters $\beta$ and $\sigma_{12}$ get
modified, with respect to the standard $\Lambda $CDM case, when CDM and
DE can interact with each other.

\subsection{The DM halo clustering in real space}
\label{xi_real}

\begin{figure*}
\includegraphics[width=\textwidth]{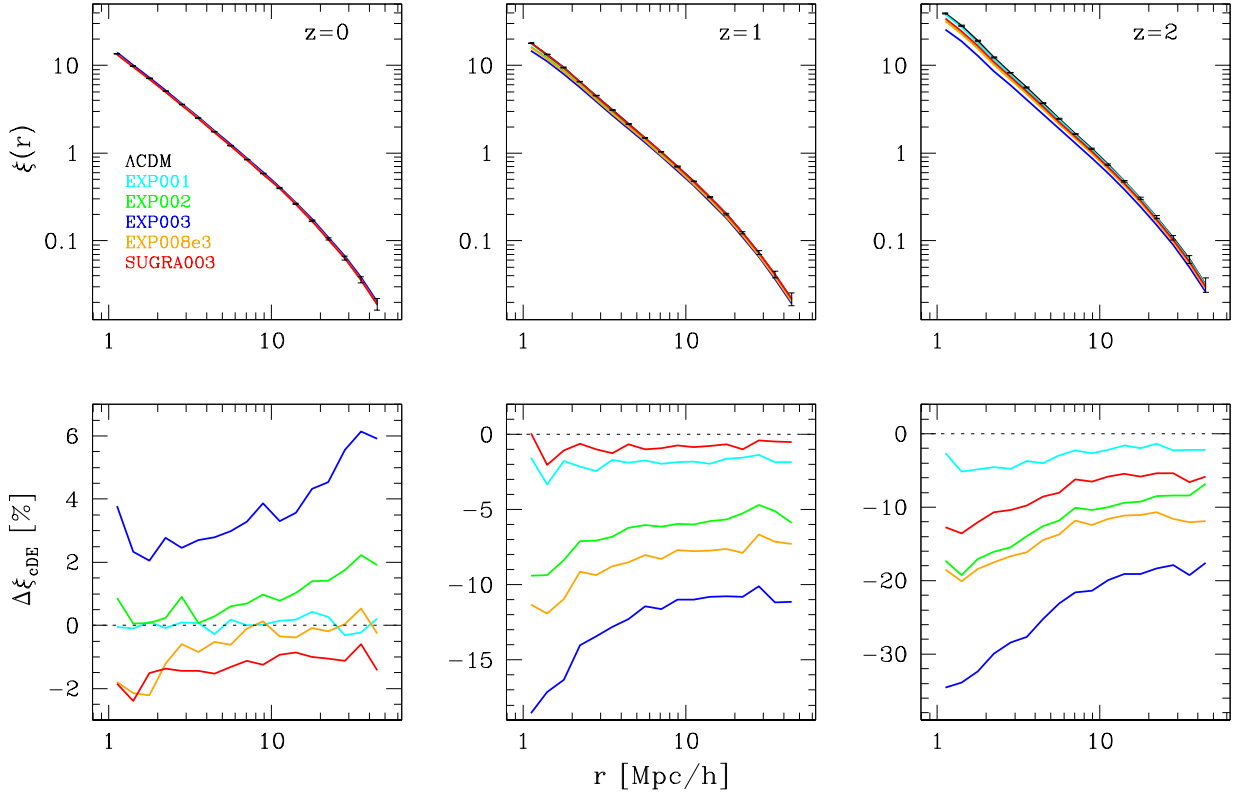}
\caption{{\em Upper panels}: real-space two-point correlation function
  of CDM haloes for all the models of the {\small CoDECS} project. The
  black error bars represent the statistical noise as prescribed by
  \citet{mo1992}. {\em Lower panels}: percentage difference between
  cDE and $\Lambda $CDM predictions, $\Delta\xi_{\rm
    cDE}=100\cdot(\xi_{\rm cDE}-\xi_{\rm \Lambda CDM})/\xi_{\rm
    \Lambda CDM}$. At $z=0$, the spatial properties of CDM haloes in
  cDE models are very similar to the $\Lambda $CDM case. However, at
  higher redshifts the DE coupling produces a significant
  scale-dependent suppression in the halo clustering.}
 \label{fig:xiReal}
\end{figure*}

In cDE scenarios the formation and evolution of cosmic structures can
differ significantly with respect to the $\Lambda $CDM case. To analyze
the spatial properties of LSS as a function of the DE coupling, we
measure the CDM halo two-point correlation function, $\xi(r)$, defined
as:
\begin{equation}
dP=n^2[1+\xi(r)]dV_1dV_2,
\end{equation}
where $dP$ is the probability of finding a halo pair with one of them
in the volume $dV_1$ and the other in the volume $dV_2$, separated by
a comoving distance $r$. To measure the function $\xi(r)$, we use the
standard \citet{landy1993} estimator:
\begin{equation}
\hat{\xi}(r)=\frac{HH(r)-2RH(r)+RR(r)}{RR(r)},
\end{equation}
where $HH(r)$, $HR(r)$ and $RR(r)$ are the fraction of halo--halo,
halo--random and random--random pairs, with spatial separation in the
range $[r-dr/2, r+dr/2]$. The random samples are three times larger
than the halo ones. A computationally efficient linked-list algorithm
has been used to count the number of pairs in the halo and random
catalogues.

Fig.~\ref{fig:xiReal} shows the real-space two-point correlation
function of CDM haloes for all the models of the {\small CoDECS}
project. The black error bars represent the statistical noise as
prescribed by \citet{mo1992}. The lower panels show the percentage
difference between cDE and $\Lambda $CDM predictions, $\Delta\xi_{\rm
  cDE}=100\cdot(\xi_{\rm cDE}-\xi_{\rm \Lambda CDM})/\xi_{\rm \Lambda
  CDM}$. The $z=0$ spatial properties of CDM haloes in cDE models are
very similar to the $\Lambda $CDM case. The EXP003 model shows the
largest deviations that are in any case quite small, between $\sim2\%$
and $\sim6\%$ for comoving separation $<50$ \Mpch. The situation is
different at higher redshifts, where the DE coupling produces a
significant scale-dependent suppression in the halo clustering. This
effect strongly depends on the coupling strength and on the
redshift. At small scales, the deviations with respect to the $\Lambda
$CDM predictions rise to $\sim20\%$ at $z=1$ and $\sim35\%$ at $z=2$
for the extreme EXP003 model. The EXP008e3 and SUGRA003 models predict
a different redshift evolution in the clustering function with respect
to the models with a constant coupling. Therefore, measuring the
redshift evolution of $\xi(r)$ can help to disentangle different
interacting DE cosmologies.


\subsection{The CDM halo biasing function}
\label{bias}

\begin{figure*}
\includegraphics[width=\textwidth]{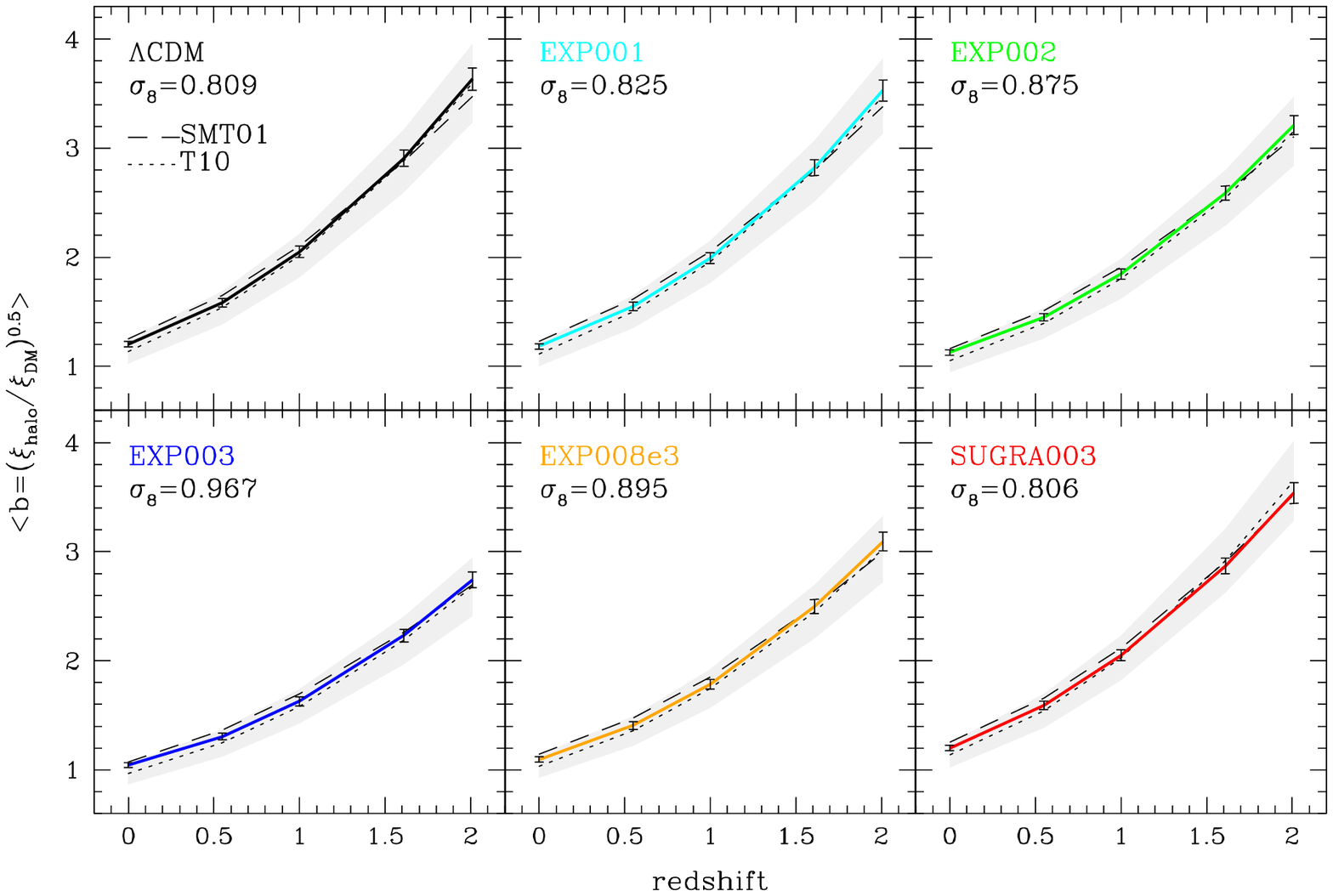}
\caption{Mean {\em apparent} effective bias of CDM haloes,
  $<b>=<(\xi_{\rm halo, cDE}/\xi_{\rm DM, \Lambda
    CDM(\sigma_{8})})^{0.5}>$, averaged in the range
  10\Mpch$<r<$50\Mpch. Dashed and dotted black lines show the
  theoretical $\Lambda $CDM effective bias computed according to the
  relations of \citet{sheth_mo_tormen2001} and \citet{tinker2010},
  respectively, normalized to the $\sigma_{8}$ values of the {\small
    CoDECS} models. The error bars represent the propagated
  statistical noise as prescribed by \citet{mo1992}, while the grey
  shaded areas show a $10\%$ error. The effect of DE coupling on the
  CDM halo large scale bias appears totally degenerate with $\sigma_{8}$
  for all the cosmological models considered.}
 \label{fig:biasz}
\end{figure*}

The clustering suppression predicted by cDE models in mass-selected
halo samples can be better understood by looking at the large-scale
halo bias. Fig.~\ref{fig:biasz} shows the mean {\em apparent} effective
bias of CDM haloes defined as follows:
\begin{equation} 
<b(z)>=<\sqrt{\frac{\xi_{\rm halo, cDE}}{\xi_{\rm DM, \Lambda CDM(\sigma_{8})}}}>,
\end{equation}
where $\xi_{\rm halo, cDE}$ is the cDE halo clustering and $\xi_{\rm
  DM, \Lambda CDM(\sigma_{8})}$ is the $\Lambda $CDM DM clustering
normalized to the $\sigma_{8}$ values of the {\small CoDECS}
models. $\xi_{\rm DM, \Lambda CDM(\sigma_{8})}$ has been obtained by
Fourier trasforming the non-linear power spectrum extracted from CAMB
\citep{lewis2002}, which exploits the HALOFIT routine
\citep{smith2003}.  The bias is averaged in the range
10\Mpch$<r<$50\Mpch. The error bars represent the propagated
statistical noise as prescribed by \citet{mo1992}. By definition,
$<b(z)>$ represents the apparent bias that would be derived in a cDE
universe if a $\Lambda $CDM model was erroneously assumed to predict
the DM clustering.

Dashed and dotted black lines in Fig.~\ref{fig:biasz} show the
theoretical $\Lambda $CDM effective bias of
\citet{sheth_mo_tormen2001} and \citet{tinker2010}, respectively,
normalized to the values of $\sigma_{8}$ of the {\small CoDECS}
models and weighted with the halo mass function, $n(M,z)$:
\begin{equation}
b_{\rm eff}(z) = \frac{\int_{M_{\rm min}}^{M_{\rm max}} n(M,z)
  b(M,z)dM}{\int_{M_{\rm min}}^{M_{\rm max}} n(M,z)dM},
\label{eq:bias} 
\end{equation}
where $M_{\rm min}$ and $M_{\rm max}$ have been defined in section \S
\ref{N-body}. The grey shaded bands show a $10\%$ error, that
approximately represents the uncertainty in the theoretical bias
predictions presented in the literature.

\begin{figure*}
\includegraphics[width=\textwidth]{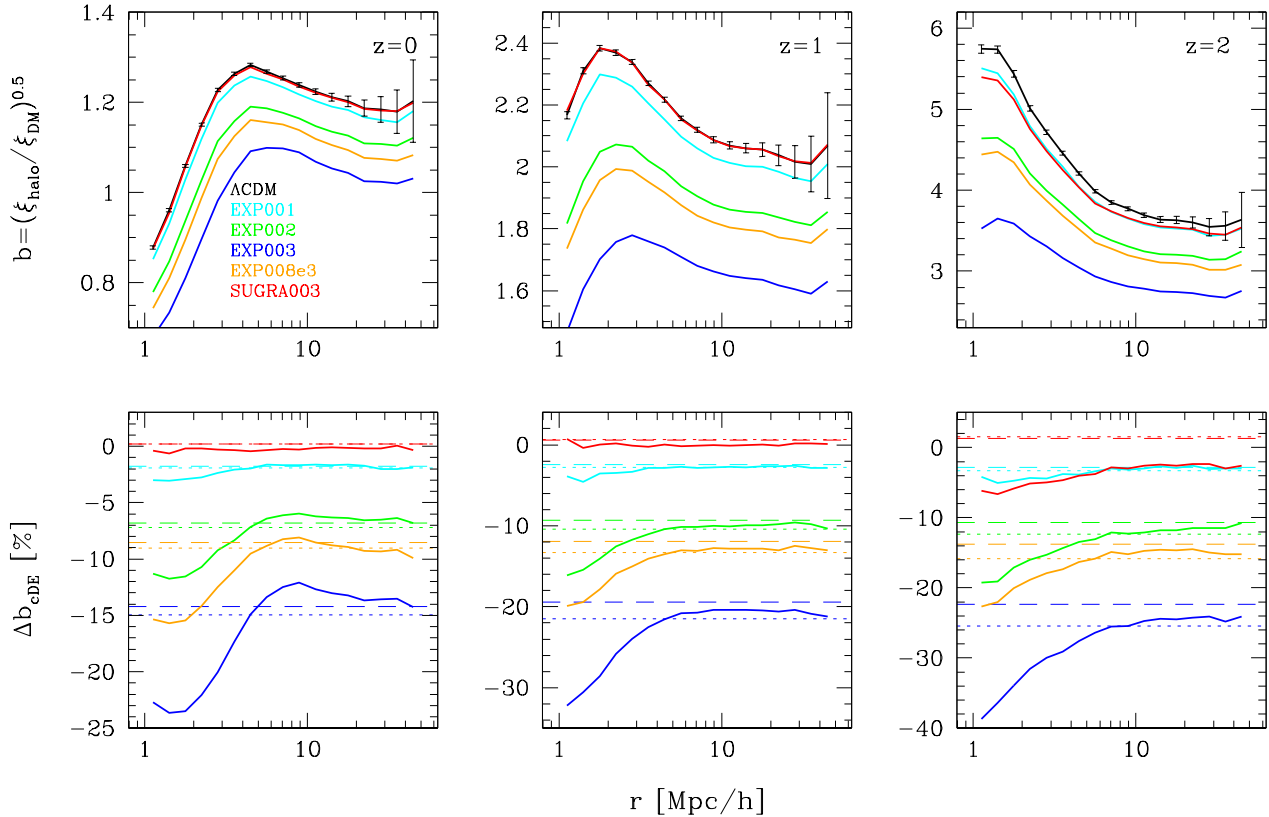}
\caption{{\em Upper panels}: {\em apparent} effective bias of CDM halo
  as a function of scale, at three different redshifts $z=0,1,2$ from
  left to right. The error bars represent the statistical noise
  \citep{mo1992}. {\em Lower panels}: percentage difference between
  cDE and $\Lambda $CDM predictions, $\Delta b_{\rm
    cDE}=100\cdot(b_{\rm cDE}-b_{\rm \Lambda CDM})/b_{\rm \Lambda
    CDM}$. Dashed and dotted lines show the theoretical $\Lambda $CDM
  predictions computed according to the relations of
  \citet{sheth_mo_tormen2001} and \citet{tinker2010}, respectively,
  normalized to the $z=0$ values of $\sigma_{8}$ of the {\small CoDECS}
  models. The scale-dependent suppression of the halo bias at small
  separations ($r\lesssim 5-10$ \Mpch) due to the DE coupling breaks
  the $\sigma_{8}$ degeneracy.}
 \label{fig:bias}
\end{figure*}

The effect of DE coupling on the CDM halo large-scale bias appears
totally degenerate with $\sigma_{8}$ for all the cosmological models
considered \citep[see also][for a general discussion of degeneracies
  in cDE models]{Clemson_etal_2011}.  As already seen in
Fig.~\ref{fig:xiReal}, the halo clustering properties at small scales
can help in removing this degeneracy. This can be better appreciated
in Fig.~\ref{fig:bias} that shows CDM halo bias as a function of
scale. The lower panels show the percentage difference between cDE and
$\Lambda $CDM predictions, $\Delta b_{\rm cDE}=100\cdot(b_{\rm
  cDE}-b_{\rm \Lambda CDM})/b_{\rm \Lambda CDM}$.  Dashed and dotted
lines refer to the theoretical $\Lambda $CDM predictions of
\citet{sheth_mo_tormen2001} and \citet{tinker2010}, respectively,
normalized to the values of $\sigma_{8}$ of the {\small CoDECS}
models. It is clear that the scale-dependent suppression of the halo
bias at small separations, due to the DE coupling, breaks the
$\sigma_{8}$ degeneracy. Indeed, at scales $r\lesssim 5-10$ \Mpch, both
the models with a constant coupling and the EXP008e3 one are not well
reproduced by a standard $\Lambda $CDM model after having rescaled the
matter power spectrum to the $z=0$ normalization of the cDE model. The
SUGRA003 appears in better agreement with the $\Lambda $CDM
predictions at $z\lesssim1$, while at $z\gtrsim2$ the halo bias of
this model is not degenerate with $\sigma_{8}$ at all scales, although
the effect is quite small ($\sim3\%$ at $z=2$).

\subsection {Redshift-space clustering distortions}
\label{z-dist}

\subsubsection{Clustering shape in redshift-space}

The redshift of a galaxy does not correspond to a unique distance from
the observer, as the line-of-sight component of a galaxy's peculiar
velocity creates an additional Doppler shift. When the distances are
computed without correcting for this peculiar velocity contribution,
we say to be in {\em redshift-space} and we refer to the
redshift-space spatial coordinates using the vector $\vec{s}$ (while
$\vec{r}$ indicates real-space coordinates). A redshift-space galaxy
map is distorted with respect to the {\em real} one. RSD provide
crucial constraints on the build-up of LSS. On small scales
($\lesssim$1 \Mpch) the distortion is mainly caused by the random
orbital motions of galaxies moving inside virialised structures,
i.e. the well-known {\em fingers of God} effect
\citep{jackson1972}. On larger distance scales, the coherent bulk
motion of virialising structures leads to an apparent excess in the
clustering strength perpendicularly to the line-of-sight.

Since both comoving coordinates and peculiar velocities are known in
an N-body simulation, we can investigate the spatial properties of LSS
both in real- and redshift-space. To obtain redshift-space halo maps,
we locate a virtual observer at $z=0$ and put each simulation box at a
comoving distance corresponding to its output redshifts. The
redshift-space distance of each CDM halo is obtained through the
following equation:
\begin{equation}
s_\parallel=\int_0^{z_{\rm obs}}\frac{c dz'_{\rm obs}}{H(z'_{\rm obs})},
\label{eq:distance}
\end{equation}
where $H(z'_{\rm obs})$ is the Hubble rate, $c$ is the speed of light and the
observed redshift, $z_{\rm obs}$, is given by:
\begin{equation}
z_{\rm obs}=z_{\rm c}+\frac{v_\parallel}{c}(1+z_{\rm c}).
\label{eq:redshift}
\end{equation}
Here $z_{\rm c}$ is the {\em cosmological} redshift, due to the Hubble
flow, while the second term of Eq.~(\ref{eq:redshift}) is caused by
galaxy peculiar velocities ($v_\parallel$ is the component of the
peculiar velocity parallel to the line-of-sight). 

Whether the cosmological parameters assumed in Eq.~(\ref{eq:distance})
are not the true ones, or the Universe is not described by the
$\Lambda $CDM equations, like in the cDE scenarios, a {\em geometric}
kind of distortion is added to the {\em dynamic} distortion caused by
peculiar velocities \citep{alcock1979}. However, as we will see in
Sec.~\ref{modelling}, geometric distortions in cDE models are rather
small relative to the dynamical ones and can therefore be neglected.

Such a simple method to derive redshift-space coordinates is not as
accurate as a full past light-cone reconstruction \citep[see
  e.g.][]{marulli2009}, since it neglects the redshift evolution of
halo properties inside each snapshot box. However, this method is
accurate enough for the purpose of the present work. In a companion
paper we aim to estimate the expected observational uncertainties in
constraining cDE model parameters with next generation galaxy surveys,
and for that specific goal realistic mock galaxy catalogues will be
constructed.

\begin{figure}
\includegraphics[width=0.45\textwidth]{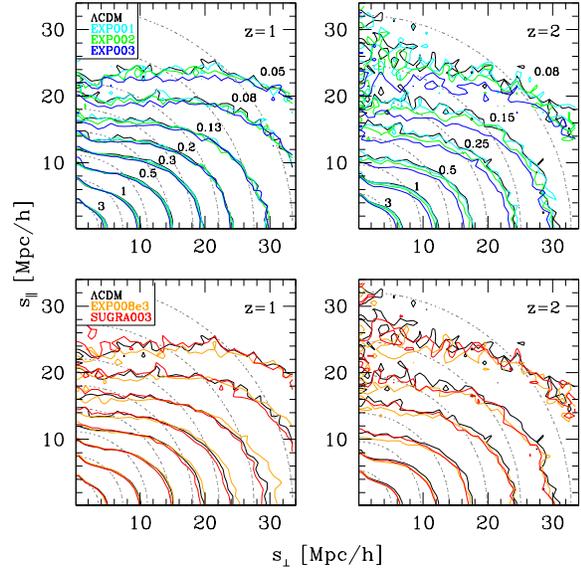}
\caption{Contour lines of the two-point redshift-space correlation function
  of CDM haloes, \xiis. The iso-curve plotted are
  \xiis$=\{0.05,0.08,0.13,0.2,0.3,0.5,1,3\}$ and
  $\{0.08,0.15,0.25,0.5,1,3\}$ in the left and right panels,
  respectively. Dotted grey lines show the undistorted correlation
  function in real-space.}
 \label{fit:iso}
\end{figure}

To extract cosmological constraints from RSD it is convenient to
decompose the distances into the two components perpendicular and
parallel to the line-of-sight, $\vec{r}=(r_\perp,r_\parallel)$. When
measured in real-space, the contour lines of \xii are circles, as the
population of haloes/galaxies is isotropic when averaged on large
scales. Instead, the redshift-space correlation function is distorted:
at small scales, the {\em fingers of God} effect changes the shape of
correlation in the direction parallel to the line-of-sight, while at
large scales the bulk motion squashes the correlation perpendicularly
to the line-of-sight. These effects can be clearly seen in
Fig.~\ref{fit:iso}, where the contour lines of the two-point
redshift-space correlation function of CDM haloes are shown. We also
plot in the same panels the undistorted real-space correlation
function (dotted grey lines) for comparison.


\subsubsection{Modelling the dynamical redshift-space distortions}
\label{modelling}

Assuming the plane-parallel approximation, the linear power spectrum of the matter
density fluctuations, $P(k)$, can be parameterized in redshift-space as follows:
\begin{equation}
P(k) = (1+\beta\mu^2)^2 P_{\rm lin}(k),
\label{eq:kaiser}
\end{equation}
where $P_{\rm lin}(k)$ is the linear power spectrum in real-space,
$\mu$ is the cosine of the angle between $\vec{k}$ and the
line-of-sight and $\beta$ is the linear distortion parameter. Fourier
transforming equation (\ref{eq:kaiser}) gives
\begin{equation} 
\xi(s,\mu)= \xi_0(s)P_0(\mu)+\xi_2(s)P_2(\mu)+\xi_4(s)P_4(\mu),
\label{eq:ximodellin}
\end{equation}
where the functions $P_l$ are the Legendre polynomials
\citep{hamilton1992}, while the multipoles, $\xi_n(s)$, can be written
as follows
\begin{equation} 
\xi_0(s) = \left(1+ \frac{2\beta}{3} + \frac{\beta^2}{5}\right)\xi(r),
\end{equation}
\begin{equation} 
\xi_2(s) = \left(\frac{4\beta}{3} +
\frac{4\beta^2}{7}\right)[\xi(r)-\overline{\xi}(r)],
\end{equation}
\begin{equation} 
\xi_4(s) = \frac{8\beta^2}{35}\left[\xi(r) +
  \frac{5}{2}\overline{\xi}(r)
  -\frac{7}{2}\overline{\overline{\xi}}(r)\right].
\end{equation}
Here $\xi(r)$ is the real-space correlation function and the {\em barred}
correlation functions are defined as
\begin{equation} 
\overline{\xi}(r) = \frac{3}{r^3}\int^r_0dr'\xi(r')r'{^2},
\end{equation}
\begin{equation}
\overline{\overline{\xi}}(r) = \frac{5}{r^5}\int^r_0dr'\xi(r')r'{^4}.
\end{equation}
To include in the model also the small non-linear scales, we convolve
it with the distribution function of random pairwise velocities,
$f(v)$ \citep[but see e.g.][]{scoccimarro2004, matsubara2004}:
\begin{equation} 
 \xi(s_\perp, s_\parallel) = \int^{\infty}_{-\infty}dv
 f(v)\xi(s_\perp, s_\parallel - v/H(z)/a(z)).
 \label{eq:ximodel}
\end{equation}
In this work, we adopt for $f(v)$ the form
\begin{equation}
f_{\rm exp}(v)=\frac{1}{\sigma_{12}\sqrt{2}}
\exp\left(-\frac{\sqrt{2}|v|}{\sigma_{12}}\right),
\label{eq:fvexp} 
\end{equation}
where $\sigma_{12}$ is the dispersion in the pairwise peculiar
velocities.

To summarize, the model for the distorted correlation function in
redshift-space (Eq.~(\ref{eq:ximodel})) depends on the correlation
function in real space, $\xi(r)$, and on two free parameters: the
linear distortion parameter, $\beta$, and the pairwise peculiar
velocities dispersion, $\sigma_{12}$. It can be demonstrated that
$\beta$ is well parameterized by the following equation:
\begin{equation}
\beta=\frac{f(\Omega_m)}{b}\simeq\frac{\Omega_m(z)^\gamma}{b},
\end{equation}
where $f(\Omega_m)=d\ln D/d\ln a$ is the velocity growth rate, $D$ is
the linear density growth factor, $b$ is the linear CDM halo bias, and
$\Omega_m(z)$ is the matter density. The so-called gravitational
growth index, $\gamma$, directly depends on the gravity theory, being
$\gamma\sim0.545$ for GR gravity. As a consequence,
observational constraints on the parameter $\beta$ that can be
obtained modelling the redshift-space correlation function \xiis
directly translate to constraints on the gravity theory.

\begin{figure*}
\includegraphics[width=\textwidth]{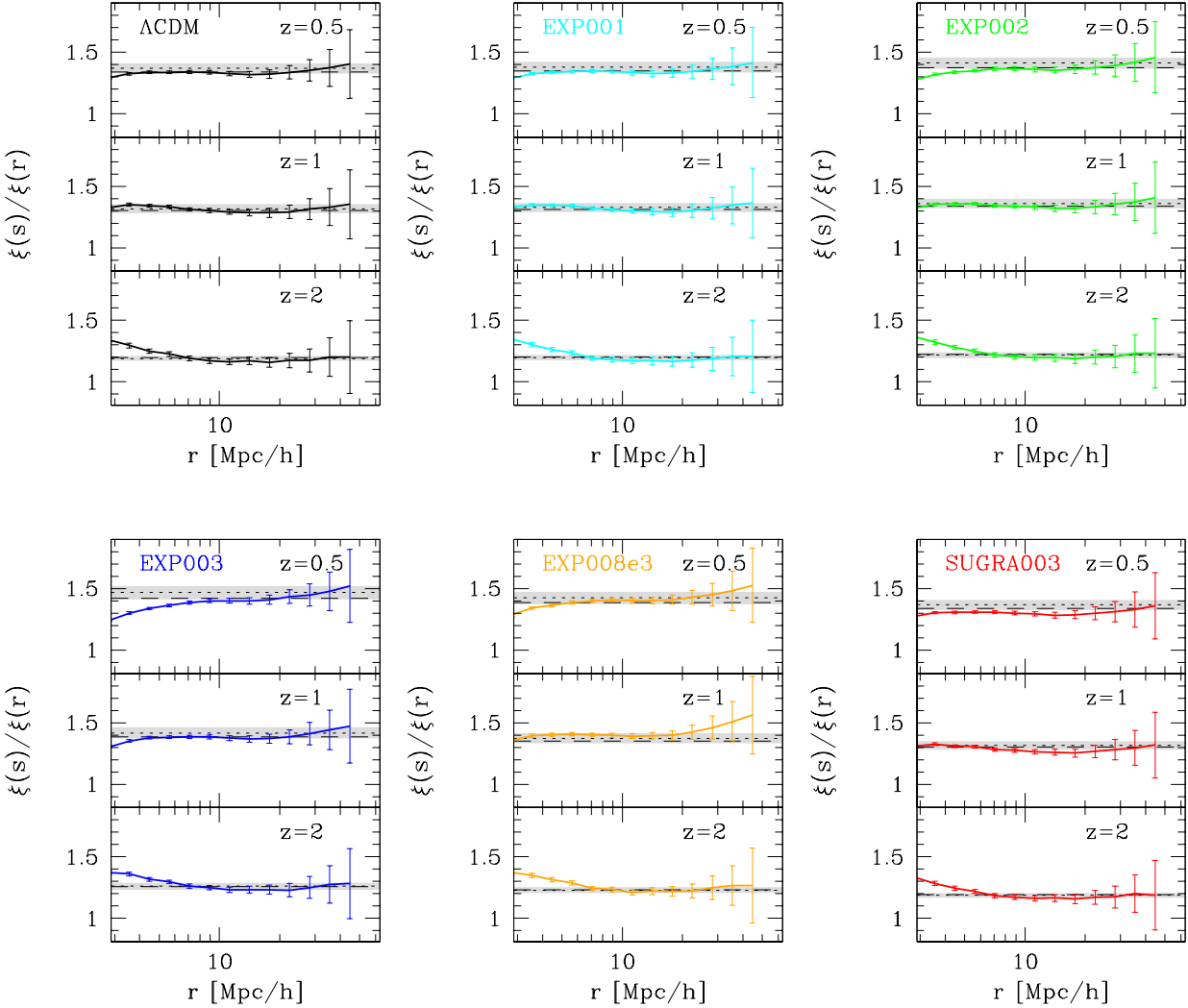}
\caption{Ratio between redshift- and real-space correlation function
  of the CDM haloes. Dashed and dotted black lines show the
  theoretical $\Lambda $CDM predictions (Eq.~(\ref{eq:xiratio})) using
  the effective bias functions computed according to the relations of
  \citet{sheth_mo_tormen2001} and \citet{tinker2010}, respectively,
  normalized to the values of $\sigma_{8}$ of the {\small CoDECS}
  models. The error bars represent the statistical noise
  \citep{mo1992}. The DE coupling strongly affects the redshift-space
  clustering distortions. However such effect is strongly degenerate
  with $\sigma_{8}$ for the redshift considered. Only the SUGRA003 model
  appears not totally degenerate with $\sigma_{8}$ at $z=0.5$, but the
  effect is small.}
\label{fig:ratio}
\end{figure*}

Fig.~\ref{fig:ratio} shows the ratio between redshift- and real-space
correlation functions of the CDM haloes in all the {\small CoDECS}
simulations. The error bars represent the statistical noise
\citep{mo1992}. The black lines show the theoretical $\Lambda $CDM
prediction given by the large-scale limit of
Eq.~(\ref{eq:ximodellin}):
\begin{equation}
\frac{\xi(s)}{\xi(r)} = 1 + \frac{2\beta}{3} + \frac{\beta^2}{5},
\label{eq:xiratio}
\end{equation}
where $\beta=\frac{\Omega_m(z)^{0.545}}{b_{\rm eff}}$, and the
effective bias $b_{\rm eff}$ is given by Eq.~(\ref{eq:bias}). Dashed
and dotted black lines refer to the theoretical $\Lambda $CDM
predictions obtained using the effective bias of
\citet{sheth_mo_tormen2001} and \citet{tinker2010}, respectively,
normalized to the values of $\sigma_{8}$ of the {\small CoDECS}
models. The grey shaded area represents the propagated $10\%$
theoretical bias error.

As expected, the measured and model ratios $\xi(s)/\xi(r)$ agree quite
well in the $\Lambda $CDM model.  The cosmic evolution of LSS in cDE
models has the effect of significantly suppressing this ratio. However,
as for the large-scale clustering and biasing function, RSD appears to
be strongly degenerate with $\sigma_{8}$ at the scales and redshifts
considered. Indeed, the mean ratio $\xi(s)/\xi(r)$ of CDM haloes in
the cDE simulation can be reproduced quite well by the expected
$\Lambda $CDM prediction rescaled at the $\sigma_{8}$ value of the cDE
model. Only the SUGRA003 model appears not totally degenerate with
$\sigma_{8}$ at $z=0.5$, but the effect is quite small.

\begin{figure}
\includegraphics[width=0.45\textwidth]{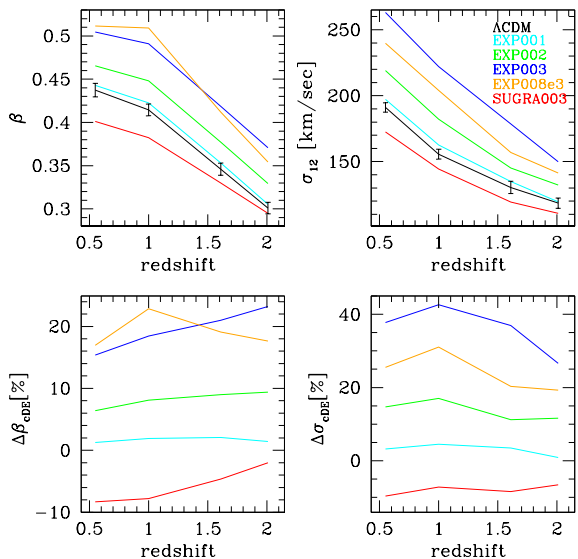}
\caption{{\em Upper panels}: Best-fit parameters $\beta(z)$ (left
  panel) and $\sigma_{12}(z)$ (right panel) measured in all the
  {\small CoDECS} simulations. The error bars shown olny for $\Lambda
  $CDM represent the scatter in the measured $\beta$ and $\sigma_{12}$
  obtained by dividing the simulation box in 27 sub-boxes, and
  rescaled by the square root of the total volume of the simulation
  box \citep[see e.g.][]{marulli2011}. {\em Lower panels}: percentage
  difference between cDE and $\Lambda $CDM predictions,
  $\Delta\beta_{\rm cDE}=100\cdot(\beta_{\rm cDE}-\beta_{\rm \Lambda
    CDM})/\beta_{\rm \Lambda CDM}$ (left panel), $\Delta\sigma_{\rm
    cDE}=100\cdot(\sigma_{12,\rm cDE}-\sigma_{12,\rm \Lambda
    CDM})/\sigma_{12,\rm \Lambda CDM}$ (right panel).}
 \label{fig:best-fit}
\end{figure}

To extract the cosmological information from RSD both at small and
large scales, we use Eq.~(\ref{eq:ximodel}) to model the
redshift-space correlation function \xiis, where the real-space
$\xi(r)$ has been measured using directly the comoving CDM halo model
coordinates. The best-fit parameters, $\beta$ and $\sigma_{12}$,
obtained with a standard $\chi^2$ minimization procedure are shown in
the upper panels of Fig.~\ref{fig:best-fit}, as a function of
redshift. The errors on the best-fit parameter predictions have been
estimated dividing the simulation box in 27 sub-boxes, measuring
$\beta$ and $\sigma_{12}$ in each of them and rescaling their scatter
by the square root of the total volume of the simulation box
\citep[see e.g.][]{marulli2011}.  The strong effect of cDE on both
these parameters can be appreciated in the lower panels that show the
percentage difference between cDE and $\Lambda $CDM predictions. Both
in the models with a constant coupling and in the EXP008e3 one,
$\beta$ and $\sigma_{12}$ are higher with respect to the $\Lambda $CDM
case. This effect depends significantly on the coupling strenght and
for the most extreme models considered raises to the level of $\sim
20\%$ for $\beta$ and $\sim 40\%$ for $\sigma_{12}$. Viceversa, the
SUGRA003 model predicts a suppression of $\sim10\%$ both in $\beta$
and in $\sigma_{12}$ at $z\sim0.5-1$, while at higher redshifts the
differences with respect to the $\Lambda $CDM model decrease.

This result suggests that using an independent constraint on
$\sigma_{8}$, that for instance can be obtained from the clustering and
bias function at small scales (see Sec.~\ref{xi_real} and \ref{bias}),
the next generation of galaxy surveys will be able to put strong
constraints on the coupling between DE and CDM exploiting the shape of
redshift-space clustering anisotropies.

\subsubsection{Geometric distortions}
\label{geometric}
\begin{figure}
\includegraphics[width=0.45\textwidth]{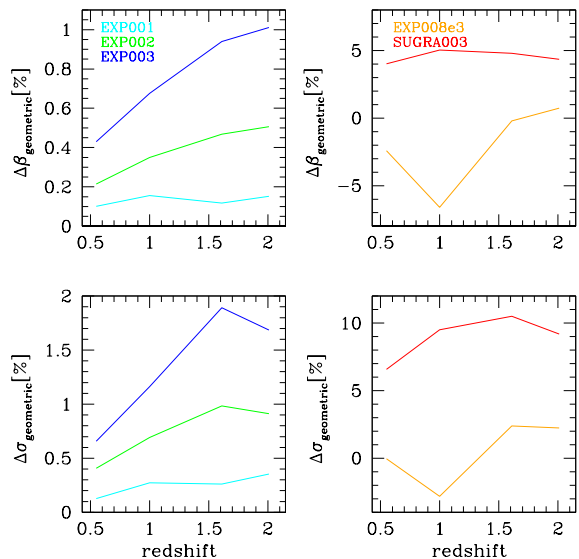}
\caption{Percentage errors on the best-fit parameters $\beta(z)$ (left
  panel) and $\sigma_{12}(z)$ (right panel) introduced if
  Eq.~(\ref{eq:distance}) is used to measure redshift-space distances
  also in the cDE simulations.}
 \label{fig:AP}
\end{figure}

In cDE scenarios, the relation between redshift and comoving distance
is different from the one in the $\Lambda $CDM case. This means that
if the $\Lambda $CDM equations are used to obtain galaxy maps from
redshift measurements in a Universe where DE and CDM interact, the
correlation function appears distorted. To investigate the effect of
such {\em geometric} distortions, also called Alcock-Paczynski
distortions \citep{alcock1979}, we have repeated all the measurements
described in the previous sections assuming the $\Lambda $CDM Hubble
parameter $H(z)$ in Eq.~(\ref{eq:distance}) to measure redshift-space
distances also in the cDE simulations. Fig.~\ref{fig:AP} shows the
percentage error on the best-fit parameters introduced in this way.
In the three models with a constant coupling (EXP001, EXP002 and
EXP003) these distortions slightly increase as a function of redshift,
but only up to the level of $\sim 1\%$ for $\beta$ and $\sim 2\%$ for
$\sigma_{12}$ at $z=2$, for the extreme EXP003 model. So, as expected,
these distortions are quite small compared to the dynamical ones.
Also in the EXP008e3 models the geometric distortions appear clearly
negligible with respect to the dynamical distortions. Only the
SUGRA003 model shows large geometric distortions, at a level of
$\sim5\%$ for $\beta$ and $\sim10\%$ for $\sigma_{12}$, comparable to
the level of dynamical distortions.  This suggests that geometric and
dynamic distortions on the redshift-space clustering of galaxies might
be strongly degenerate for the ``Bouncing" cDE scenario, and
independent measurements as \eg ~the redshift evolution of the halo
mass function should be used to break the degeneracy.

\section{Conclusions} 
\label{concl}

In the present paper we have investigated the impact of a possible
interaction between DE and CDM on the clustering and redshift-space
distortions of CDM haloes, also in view of the large wealth of
high-quality data expected from the next generation of large galaxy
surveys.

In cDE scenarios, in fact, the formation and evolution of LSS can be
significantly different with respect to the $\Lambda $CDM case.  The
interaction induces a time evolution of the mass of CDM particles as
well as a modification of the growth of structures determined by the
combined effect of a long-range fifth-force mediated by the DE scalar
degree of freedom and of an extra-friction arising as a consequence of
momentum conservation.  The combination of these effects can
significantly alter the evolution of linear and nonlinear density
perturbations as compared to the standard $\Lambda $CDM scenario, and
provide a direct way to test this class of cosmological models.  In
particular, the clustering properties of CDM haloes from linear to
highly nonlinear scales might show specific signatures allowing to
distinguish a cDE universe from $\Lambda $CDM.

In order to investigate the spatial properties of LSS in these
cosmological scenarios it is therefore necessary to rely on a fully
nonlinear treatment of structure formation in the context of cDE
models.  To this end, we made use of the public halo catalogues
extracted from the largest set of N-body experiments to date for such
cosmologies, the {\small CoDECS} simulations.  From such catalogues,
we measured the clustering and bias properties of CDM haloes both in
real- and redshift-space, as a function of the DE coupling and
redshift. Moreover, we investigated the effects of the DE coupling on
the geometric and dynamic redshift-space distortions, quantifying the
difference with respect to the concordance $\Lambda $CDM model.  We
compared numerical predictions with theoretical expectations finding a
good agreement in $\Lambda $CDM, as expected, and a strong degeneracy
between the DE coupling and the matter power spectrum normalization,
$\sigma_{8}$.

In particular, at $z\sim 0$ the spatial properties of CDM haloes in
cDE models appear very similar to the $\Lambda $CDM case, even if the
{\small CoDECS} simulations are normalized to be consistent with the
latest CMB data at last scattering.  At higher redshifts instead, we
find that the DE coupling produces a significant scale-dependent
suppression in the halo clustering and bias function. Furthermore, the
DE coupling strongly affects the redshift-space clustering
distortions. We quantified this effect by modelling RSD in terms of the
redshift-space correlation function and the distribution function of
random pairwise velocities and deriving the linear distortion
parameter, $\beta$, and the pairwise peculiar velocity dispersion,
$\sigma_{12}$, for all the {\small CoDECS} models. We find that for
cDE models characterized by an exponential self-interaction potential,
both a constant coupling and an exponentially growing coupling
determine higher values of $\beta$ and $\sigma_{12}$ with respect to
the $\Lambda $CDM case. On the contrary, Bouncing cDE models based on
a SUGRA self-interaction potential predict a suppression of $\beta$
and $\sigma_{12}$ at $z\sim0.5-1$, while at higher redshifts the
differences with respect to the $\Lambda $CDM model decrease.

We also analyzed the geometric distortions induced by the DE coupling,
finding that these are always negligible with respect to dynamical
ones for all the exponential potential models, while they might have a
comparable effect as the latter for the Bouncing cDE scenario.

Overall, we found a strong degeneracy of the DE coupling phenomenology
with the amplitude of linear density perturbations defined by
$\sigma_{8}$: the clustering, the bias function and the RSD in cDE
scenarios can be well reproduced by a standard $\Lambda $CDM model
after having rescaled the matter power spectrum to the $\sigma_{8}$
normalization of each cDE model, for scales larger than $\sim5-10$
\Mpch. However, we also find that the $\sigma_{8}$ degeneracy is broken
at smaller scales.  Our results therefore suggest that it is possible
to put strong constraints on the coupling between DE and CDM
exploiting the shape of redshift-space clustering anisotropies and
comparing the clustering and bias functions at small and large scales.
\ \\

To conclude, we performed an extended analysis of the impact of an
interaction between DE and CDM on the clustering and redshift-space
distortions properties of CDM haloes, using state-of-the-art N-body
simulations for these cosmologies.  Although the effect of the
coupling is highly degenerate with $\sigma_{8}$ at large scales, at
smaller scales the degeneracy is broken, offering the chance that the
next generation of wide and high-precision measurements of the galaxy
distribution in real- and redshift-space might be able to tightly
constrain or possibly detect a new interaction in the dark sector.

\section*{acknowledgments}
We warmly thank C. Carbone and E. Branchini for helpful discussions
and suggestions. We acknowledge financial contributions from
contracts ASI-INAF I/023/05/0, ASI-INAF I/088/06/0, ASI I/016/07/0
’COFIS’, ASI ’Euclid-DUNE’ I/064/08/0, ASI-Uni Bologna-Astronomy
Dept. ’Euclid-NIS’ I/039/10/0, and PRIN MIUR ’Dark energy and
cosmology with large galaxy surveys’.\\ MB is supported by the DFG
Cluster of Excellence ``Origin and Structure of the Universe'' and by
the TRR33 Transregio Collaborative Research Network on the ``Dark
Universe''.  MB also acknowledges the HPC-Europa2 visiting programme
for financial support during his visits to the Astronomy Department in
Bologna.

\bibliographystyle{mn2e} \bibliography{bib_extended}

\begin{thebibliography}{}

\bibitem[\protect\citeauthoryear{{Alcock} \& {Paczynski}}{{Alcock} \&
  {Paczynski}}{1979}]{alcock1979}
{Alcock} C.,  {Paczynski} B.,  1979, \nat, 281, 358

\bibitem[\protect\citeauthoryear{{Allen}, {Schmidt}, {Ebeling}, {Fabian} \&
  {van Speybroeck}}{{Allen} et~al.}{2004}]{allen2004}
{Allen} S.~W.,  {Schmidt} R.~W.,  {Ebeling} H.,  {Fabian} A.~C.,    {van
  Speybroeck} L.,  2004, \mnras, 353, 457

\bibitem[\protect\citeauthoryear{{Amendola}}{{Amendola}}{2000}]{amendola2000}
{Amendola} L.,  2000, \prd, 62, 043511

\bibitem[\protect\citeauthoryear{{Amendola}}{{Amendola}}{2004}]{amendola2004}
{Amendola} L.,  2004, \prd, 69, 103524

\bibitem[\protect\citeauthoryear{{Amendola}, {Baldi} \& {Wetterich}}{{Amendola}
  et~al.}{2008}]{amendola2008}
{Amendola} L.,  {Baldi} M.,    {Wetterich} C.,  2008, \prd, 78, 023015

\bibitem[\protect\citeauthoryear{Astier et~al.,}{Astier  et~al.}{2006}]{SNLS}
Astier P.,  et~al., 2006, Astron. Astrophys., 447, 31

\bibitem[\protect\citeauthoryear{{Astier} et~al.,}{{Astier}
  et~al.}{2006}]{astier2006}
{Astier} P.,  et~al., 2006, \aap, 447, 31

\bibitem[\protect\citeauthoryear{{Baldi}}{{Baldi}}{2011a}]{Baldi_2011b}
{Baldi} M.,  2011a, Mon. Not. Roy. Astron. Soc., 414, 116

\bibitem[\protect\citeauthoryear{{Baldi}}{{Baldi}}{2011b}]{Baldi_2011c}
{Baldi} M.,  2011b, Mon. Not. Roy. Astron. Soc. submitted [arXiv:1107.5049]

\bibitem[\protect\citeauthoryear{{Baldi}}{{Baldi}}{2011c}]{CoDECS}
{Baldi} M.,  2011c, arXiv:1109.5695

\bibitem[\protect\citeauthoryear{{Baldi}}{{Baldi}}{2011d}]{Baldi_2011a}
{Baldi} M.,  2011d, Mon. Not. Roy. Astron. Soc., 411, 1077

\bibitem[\protect\citeauthoryear{{Baldi} \& {Pettorino}}{{Baldi} \&
  {Pettorino}}{2011}]{Baldi_Pettorino_2011}
{Baldi} M.,  {Pettorino} V.,  2011, Mon. Not. Roy. Astron. Soc., 412, L1

\bibitem[\protect\citeauthoryear{{Baldi}, {Pettorino}, {Robbers} \&
  {Springel}}{{Baldi} et~al.}{2010}]{Baldi_etal_2010}
{Baldi} M.,  {Pettorino} V.,  {Robbers} G.,    {Springel} V.,  2010, Mon. Not.
  Roy. Astron. Soc., 403, 1684

\bibitem[\protect\citeauthoryear{{Baugh}, {Gaztanaga} \& {Efstathiou}}{{Baugh}
  et~al.}{1995}]{baugh1995}
{Baugh} C.~M.,  {Gaztanaga} E.,    {Efstathiou} G.,  1995, \mnras, 274, 1049

\bibitem[\protect\citeauthoryear{{Binney} \& {Evans}}{{Binney} \&
  {Evans}}{2001}]{binney2001}
{Binney} J.~J.,  {Evans} N.~W.,  2001, \mnras, 327, L27

\bibitem[\protect\citeauthoryear{{Brax} \& {Martin}}{{Brax} \&
  {Martin}}{1999}]{brax1999}
{Brax} P.~H.,  {Martin} J.,  1999, Physics Letters B, 468, 40

\bibitem[\protect\citeauthoryear{{Cabr{\'e}} \& {Gazta{\~n}aga}}{{Cabr{\'e}} \&
  {Gazta{\~n}aga}}{2009a}]{cabre2009}
{Cabr{\'e}} A.,  {Gazta{\~n}aga} E.,  2009a, \mnras, 393, 1183

\bibitem[\protect\citeauthoryear{{Cabr{\'e}} \& {Gazta{\~n}aga}}{{Cabr{\'e}} \&
  {Gazta{\~n}aga}}{2009b}]{cabre2009b}
{Cabr{\'e}} A.,  {Gazta{\~n}aga} E.,  2009b, \mnras, 396, 1119

\bibitem[\protect\citeauthoryear{{Carnero}, {Sanchez}, {Crocce}, {Cabre} \&
  {Gaztanaga}}{{Carnero} et~al.}{2011}]{carnero2011}
{Carnero} A.,  {Sanchez} E.,  {Crocce} M.,  {Cabre} A.,    {Gaztanaga} E.,
  2011, ArXiv e-prints

\bibitem[\protect\citeauthoryear{Clemson, Koyama, Zhao, Maartens \&
  Valiviita}{Clemson et~al.}{2011}]{Clemson_etal_2011}
Clemson T.,  Koyama K.,  Zhao G.-B.,  Maartens R.,    Valiviita J.,  2011,
  arXiv:1109.6234

\bibitem[\protect\citeauthoryear{{Crocce}, {Gaztanaga}, {Cabre}, {Carnero} \&
  {Sanchez}}{{Crocce} et~al.}{2011}]{crocce2011}
{Crocce} M.,  {Gaztanaga} E.,  {Cabre} A.,  {Carnero} A.,    {Sanchez} E.,
  2011, ArXiv e-prints

\bibitem[\protect\citeauthoryear{{Davis}, {Efstathiou}, {Frenk} \&
  {White}}{{Davis} et~al.}{1985}]{davis1985}
{Davis} M.,  {Efstathiou} G.,  {Frenk} C.~S.,    {White} S.~D.~M.,  1985, \apj,
  292, 371

\bibitem[\protect\citeauthoryear{{Guzzo} et~al.,}{{Guzzo}
  et~al.}{2008}]{guzzo2008}
{Guzzo} L.,  et~al., 2008, \nat, 451, 541

\bibitem[\protect\citeauthoryear{{Hamilton}}{{Hamilton}}{1992}]{hamilton1992}
{Hamilton} A.~J.~S.,  1992, \apjl, 385, L5

\bibitem[\protect\citeauthoryear{{Hawkins} et~al.,}{{Hawkins}
  et~al.}{2003}]{hawkins2003}
{Hawkins} E.,  et~al., 2003, \mnras, 346, 78

\bibitem[\protect\citeauthoryear{Honorez, Reid, Mena, Verde \& Jimenez}{Honorez
  et~al.}{2010}]{Honorez_etal_2010}
Honorez L.~L.,  Reid B.~A.,  Mena O.,  Verde L.,    Jimenez R.,  2010, JCAP,
  1009, 029

\bibitem[\protect\citeauthoryear{{Jackson}}{{Jackson}}{1972}]{jackson1972}
{Jackson} J.~C.,  1972, \mnras, 156, 1P

\bibitem[\protect\citeauthoryear{{Jee}, {Rosati}, {Ford}, {Dawson}, {Lidman},
  {Perlmutter}, {Demarco}, {Strazzullo}, {Mullis}, {B{\"o}hringer} \&
  {Fassbender}}{{Jee} et~al.}{2009}]{jee2009}
{Jee} M.~J.,  {Rosati} P.,  {Ford} H.~C.,  {Dawson} K.~S.,  {Lidman} C.,
  {Perlmutter} S.,  {Demarco} R.,  {Strazzullo} V.,  {Mullis} C.,
  {B{\"o}hringer} H.,    {Fassbender} R.,  2009, \apj, 704, 672

\bibitem[\protect\citeauthoryear{{Komatsu} et~al.,}{{Komatsu}
  et~al.}{2009}]{komatsu2010}
{Komatsu} E.,  et~al., 2009, \apjs, 180, 330

\bibitem[\protect\citeauthoryear{{Komatsu} et~al.,}{{Komatsu}
  et~al.}{2011}]{komatsu2011}
{Komatsu} E.,  et~al., 2011, \apjs, 192, 18

\bibitem[\protect\citeauthoryear{Kowalski et~al.,}{Kowalski
  et~al.}{2008}]{Kowalski_etal_2008}
Kowalski M.,  et~al., 2008, Astrophys. J., 686, 749

\bibitem[\protect\citeauthoryear{Koyama, Maartens \& Song}{Koyama
  et~al.}{2009}]{Koyama_etal_2009}
Koyama K.,  Maartens R.,    Song Y.-S.,  2009, JCAP, 0910, 017

\bibitem[\protect\citeauthoryear{{Landy} \& {Szalay}}{{Landy} \&
  {Szalay}}{1993}]{landy1993}
{Landy} S.~D.,  {Szalay} A.~S.,  1993, \apj, 412, 64

\bibitem[\protect\citeauthoryear{{LaRoque}, {Bonamente}, {Carlstrom}, {Joy},
  {Nagai}, {Reese} \& {Dawson}}{{LaRoque} et~al.}{2006}]{laroque2006}
{LaRoque} S.~J.,  {Bonamente} M.,  {Carlstrom} J.~E.,  {Joy} M.~K.,  {Nagai}
  D.,  {Reese} E.~D.,    {Dawson} K.~S.,  2006, \apj, 652, 917

\bibitem[\protect\citeauthoryear{{Laureijs}}{{Laureijs}}{2009}]{laureijs2009}
{Laureijs} R.,  2009, ArXiv e-prints

\bibitem[\protect\citeauthoryear{Lee \& Baldi}{Lee \&
  Baldi}{2011}]{Lee_Baldi_2011}
Lee J.,  Baldi M.,  2011, arXiv:1110.0015

\bibitem[\protect\citeauthoryear{{Lewis} \& {Bridle}}{{Lewis} \&
  {Bridle}}{2002}]{lewis2002}
{Lewis} A.,  {Bridle} S.,  2002, \prd, 66, 103511

\bibitem[\protect\citeauthoryear{{Linder}}{{Linder}}{2008}]{linder2008}
{Linder} E.~V.,  2008, Astroparticle Physics, 29, 336

\bibitem[\protect\citeauthoryear{{Lucchin} \& {Matarrese}}{{Lucchin} \&
  {Matarrese}}{1985}]{lucchin1985}
{Lucchin} F.,  {Matarrese} S.,  1985, \prd, 32, 1316

\bibitem[\protect\citeauthoryear{{Marulli}, {Bonoli}, {Branchini}, {Gilli},
  {Moscardini} \& {Springel}}{{Marulli} et~al.}{2009}]{marulli2009}
{Marulli} F.,  {Bonoli} S.,  {Branchini} E.,  {Gilli} R.,  {Moscardini} L.,
  {Springel} V.,  2009, \mnras, 396, 1404

\bibitem[\protect\citeauthoryear{{Marulli}, {Carbone}, {Viel}, {Moscardini} \&
  {Cimatti}}{{Marulli} et~al.}{2011}]{marulli2011}
{Marulli} F.,  {Carbone} C.,  {Viel} M.,  {Moscardini} L.,    {Cimatti} A.,
  2011, \mnras, 418, 346

\bibitem[\protect\citeauthoryear{{Matsubara}}{{Matsubara}}{2004}]{matsubara200%
4}
{Matsubara} T.,  2004, \apj, 615, 573

\bibitem[\protect\citeauthoryear{{Mo}, {Jing} \& {Boerner}}{{Mo}
  et~al.}{1992}]{mo1992}
{Mo} H.~J.,  {Jing} Y.~P.,    {Boerner} G.,  1992, \apj, 392, 452

\bibitem[\protect\citeauthoryear{{Mortonson}, {Hu} \& {Huterer}}{{Mortonson}
  et~al.}{2011}]{mortonson2011}
{Mortonson} M.~J.,  {Hu} W.,    {Huterer} D.,  2011, \prd, 83, 023015

\bibitem[\protect\citeauthoryear{{Newman}, {Treu}, {Ellis}, {Sand}, {Richard},
  {Marshall}, {Capak} \& {Miyazaki}}{{Newman} et~al.}{2009}]{newman2009}
{Newman} A.~B.,  {Treu} T.,  {Ellis} R.~S.,  {Sand} D.~J.,  {Richard} J.,
  {Marshall} P.~J.,  {Capak} P.,    {Miyazaki} S.,  2009, \apj, 706, 1078

\bibitem[\protect\citeauthoryear{{Okumura}, {Matsubara}, {Eisenstein}, {Kayo},
  {Hikage}, {Szalay} \& {Schneider}}{{Okumura} et~al.}{2008}]{okumura2008}
{Okumura} T.,  {Matsubara} T.,  {Eisenstein} D.~J.,  {Kayo} I.,  {Hikage} C.,
  {Szalay} A.~S.,    {Schneider} D.~P.,  2008, \apj, 676, 889

\bibitem[\protect\citeauthoryear{{Peacock} et~al.,}{{Peacock}
  et~al.}{2001}]{peacock2001}
{Peacock} J.~A.,  et~al., 2001, \nat, 410, 169

\bibitem[\protect\citeauthoryear{{Percival} et~al.,}{{Percival}
  et~al.}{2001}]{percival2001}
{Percival} W.~J.,  et~al., 2001, \mnras, 327, 1297

\bibitem[\protect\citeauthoryear{{Percival} et~al.,}{{Percival}
  et~al.}{2004}]{percival2004}
{Percival} W.~J.,  et~al., 2004, \mnras, 353, 1201

\bibitem[\protect\citeauthoryear{{Perlmutter} et~al.,}{{Perlmutter}
  et~al.}{1999}]{perlmutter1999}
{Perlmutter} S.,  et~al., 1999, \apj, 517, 565

\bibitem[\protect\citeauthoryear{{Pettorino} \& {Baccigalupi}}{{Pettorino} \&
  {Baccigalupi}}{2008}]{pettorino2008}
{Pettorino} V.,  {Baccigalupi} C.,  2008, \prd, 77, 103003

\bibitem[\protect\citeauthoryear{Refregier et~al.,}{Refregier
  et~al.}{2010}]{EUCLID-y}
Refregier A.,  et~al., 2010, arXiv:1001.0061

\bibitem[\protect\citeauthoryear{{Riess} et~al.,}{{Riess}
  et~al.}{1998}]{riess1998}
{Riess} A.~G.,  et~al., 1998, \aj, 116, 1009

\bibitem[\protect\citeauthoryear{{Rosati} et~al.,}{{Rosati}
  et~al.}{2009}]{rosati2009}
{Rosati} P.,  et~al., 2009, \aap, 508, 583

\bibitem[\protect\citeauthoryear{{Schlegel} et~al.,}{{Schlegel}
  et~al.}{2009}]{schlegel2009}
{Schlegel} D.~J.,  et~al., 2009, ArXiv e-prints

\bibitem[\protect\citeauthoryear{Schmidt et~al.,}{Schmidt
  et~al.}{1998}]{Schmidt_etal_1998}
Schmidt B.~P.,  et~al., 1998, Astrophys.J., 507, 46

\bibitem[\protect\citeauthoryear{{Scoccimarro}}{{Scoccimarro}}{2004}]{scoccima%
rro2004}
{Scoccimarro} R.,  2004, \prd, 70, 083007

\bibitem[\protect\citeauthoryear{{Sherwin} et~al.,}{{Sherwin}
  et~al.}{2011}]{sherwin2011}
{Sherwin} B.~D.,  et~al., 2011, ArXiv e-prints

\bibitem[\protect\citeauthoryear{{Sheth}, {Mo} \& {Tormen}}{{Sheth}
  et~al.}{2001}]{sheth_mo_tormen2001}
{Sheth} R.~K.,  {Mo} H.~J.,    {Tormen} G.,  2001, \mnras, 323, 1

\bibitem[\protect\citeauthoryear{{Simon}, {Bolatto}, {Leroy} \&
  {Blitz}}{{Simon} et~al.}{2003}]{simon2003}
{Simon} J.~D.,  {Bolatto} A.~D.,  {Leroy} A.,    {Blitz} L.,  2003, \apj, 596,
  957

\bibitem[\protect\citeauthoryear{{Smith}, {Peacock}, {Jenkins}, {White},
  {Frenk}, {Pearce}, {Thomas}, {Efstathiou} \& {Couchman}}{{Smith}
  et~al.}{2003}]{smith2003}
{Smith} R.~E.,  {Peacock} J.~A.,  {Jenkins} A.,  {White} S.~D.~M.,  {Frenk}
  C.~S.,  {Pearce} F.~R.,  {Thomas} P.~A.,  {Efstathiou} G.,    {Couchman}
  H.~M.~P.,  2003, \mnras, 341, 1311

\bibitem[\protect\citeauthoryear{{Spergel} et~al.,}{{Spergel}
  et~al.}{2003}]{spergel2003}
{Spergel} D.~N.,  et~al., 2003, \apjs, 148, 175

\bibitem[\protect\citeauthoryear{{Springel}}{{Springel}}{2005}]{springel2005ga%
dget2}
{Springel} V.,  2005, \mnras, 364, 1105

\bibitem[\protect\citeauthoryear{{Springel}, {Yoshida} \& {White}}{{Springel}
  et~al.}{2001}]{springel2001}
{Springel} V.,  {Yoshida} N.,    {White} S.~D.~M.,  2001, New Astronomy, 6, 79

\bibitem[\protect\citeauthoryear{{Tegmark} et~al.,}{{Tegmark}
  et~al.}{2004}]{tegmark2004}
{Tegmark} M.,  et~al., 2004, \apj, 606, 702

\bibitem[\protect\citeauthoryear{{Tegmark} et~al.,}{{Tegmark}
  et~al.}{2006}]{tegmark2006}
{Tegmark} M.,  et~al., 2006, \prd, 74, 123507

\bibitem[\protect\citeauthoryear{{Tinker}, {Robertson}, {Kravtsov}, {Klypin},
  {Warren}, {Yepes} \& {Gottl{\"o}ber}}{{Tinker} et~al.}{2010}]{tinker2010}
{Tinker} J.~L.,  {Robertson} B.~E.,  {Kravtsov} A.~V.,  {Klypin} A.,  {Warren}
  M.~S.,  {Yepes} G.,    {Gottl{\"o}ber} S.,  2010, \apj, 724, 878

\bibitem[\protect\citeauthoryear{{Ursino}, {Branchini}, {Galeazzi}, {Marulli},
  {Moscardini}, {Piro}, {Roncarelli} \& {Takei}}{{Ursino}
  et~al.}{2011}]{ursino2011}
{Ursino} E.,  {Branchini} E.,  {Galeazzi} M.,  {Marulli} F.,  {Moscardini} L.,
  {Piro} L.,  {Roncarelli} M.,    {Takei} Y.,  2011, \mnras, 414, 2970

\bibitem[\protect\citeauthoryear{{Vikhlinin}, {Kravtsov}, {Forman}, {Jones},
  {Markevitch}, {Murray} \& {Van Speybroeck}}{{Vikhlinin}
  et~al.}{2006}]{vikhlinin2006}
{Vikhlinin} A.,  {Kravtsov} A.,  {Forman} W.,  {Jones} C.,  {Markevitch} M.,
  {Murray} S.~S.,    {Van Speybroeck} L.,  2006, \apj, 640, 691

\bibitem[\protect\citeauthoryear{{Waizmann}, {Ettori} \&
  {Moscardini}}{{Waizmann} et~al.}{2011}]{waizmann2011}
{Waizmann} J.-C.,  {Ettori} S.,    {Moscardini} L.,  2011, \mnras, pp 1415--+

\bibitem[\protect\citeauthoryear{{Wetterich}}{{Wetterich}}{1988}]{wetterich198%
8}
{Wetterich} C.,  1988, Nuclear Physics B, 302, 668

\bibitem[\protect\citeauthoryear{{Wetterich}}{{Wetterich}}{1995}]{wetterich199%
5}
{Wetterich} C.,  1995, \aap, 301, 321

\bibitem[\protect\citeauthoryear{{White}}{{White}}{1994}]{white1994}
{White} S.~D.~M.,  1994, ArXiv Astrophysics e-prints

\bibitem[\protect\citeauthoryear{{Zehavi}, {Eisenstein}, {Nichol}, {Blanton},
  {Hogg}, {Brinkmann}, {Loveday}, {Meiksin}, {Schneider} \& {Tegmark}}{{Zehavi}
  et~al.}{2005}]{zehavi2005}
{Zehavi} I.,  {Eisenstein} D.~J.,  {Nichol} R.~C.,  {Blanton} M.~R.,  {Hogg}
  D.~W.,  {Brinkmann} J.,  {Loveday} J.,  {Meiksin} A.,  {Schneider} D.~P.,
  {Tegmark} M.,  2005, \apj, 621, 22

\end{thebibliography}

\end{document}